\documentclass[12pt,preprint]{aastex}
\usepackage{natbib}
\pdfoutput=1
\usepackage{subfigure}
\usepackage{amssymb,amsmath}

\shorttitle{Magellanic Bridge}
\shortauthors{NOEL ET AL.}

\begin{document}


\title{The MAGellanic Inter-Cloud project (MAGIC) I: Evidence for intermediate-age stellar populations in between the Magellanic Clouds}

%

\author{N. E. D. No\"{e}l\altaffilmark{1,}\altaffilmark{2,5}, B. C. Conn\altaffilmark{2}, R. Carrera\altaffilmark{3,}\altaffilmark{4}, J. I. Read\altaffilmark{1,5}, H.-W. Rix\altaffilmark{2} and A. Dolphin\altaffilmark{6}
}
\altaffiltext{1}{ETH Z\"{u}rich, Institute for Astronomy, Wolfgang-Pauli-Strasse 27, 8093 Z\"urich, Switzerland. E-mail: noelia@phys.ethz.ch}
\altaffiltext{2}{Max Planck Institut f\"{u}r Astronomie, K\"{o}nigstuhl 17, 69117, Heidelberg, Germany}
\altaffiltext{3}{Instituto de Astrof{\'i}sica de Canarias, C/ V{\'i}a L\'actea s/n, 38200, La Laguna, Tenerife, Spain}
\altaffiltext{4}{Departamento de Astrof{\'i}sica, Universidad de La Laguna, Tenerife, Spain}
\altaffiltext{5}{Department of Physics, University of Surrey, Guildford, GU2 7XH, UK}
\altaffiltext{6}{Raytheon Company, PO Box 11337, Tucson, AZ, 85734-1337, United States of America}
  



\begin{abstract}
The origin of the gas in between the Magellanic Clouds (MCs) -- known as the `Magellanic Bridge' (MB) -- is puzzling. Numerical simulations suggest that the MB formed from tidally stripped gas and stars in a recent interaction between the MCs. However, the apparent lack of stripped intermediate- or old-age stars associated with the MB is at odds with this picture. In this paper, we present the first results from the MAGellanic Inter-Cloud program (MAGIC) aimed at probing the stellar populations in the inter-cloud region. We present observations of the stellar populations in two large fields located in between the Large and Small Magellanic Clouds (LMC/SMC), secured using the WFI camera on the 2.2 m telescope in La Silla. Using a synthetic color-magnitude diagram (CMD) technique, we present the first quantitative evidence for the presence of intermediate-age and old stars in the inter-Cloud region. The intermediate-age stars -- that make up $\sim$28\% of all stars in the region -- are not present in fields at a similar distance from the SMC in a direction pointing away from the LMC. This provides potential evidence that these intermediate-age stars could have been tidally stripped from the SMC. However, spectroscopic studies will be needed to confirm or rule out the tidal origin for the inter-Cloud gas and stars.
\end{abstract}


\section{Introduction}
Encounters between galaxies, common in the early Universe, are still frequent today (e.g. \citealt{1978ApJ...219...46L}, \citealt{1992Natur.360..715B}). Interactions reshape the galaxies, transferring mass, energy, and metals between them, creating new sites of star formation both within the galaxies and in stripped gas between the galaxies. As such, understanding these encounters is key to understanding galaxy formation and evolution. 

One of the best places for studying such interactions is the Local Group, where galaxies are close enough that they can be resolved into individual stars. In particular, the Magellanic Clouds (MCs), are ideal for investigating the effects of interactions due to three key factors: their close proximity to our Galaxy ($\sim $50-70\,kpc away); their mutual association over the past few Gyrs (as determined from their proper motions: 
\citealt{2006ApJ...652.1213K}, \citealt{2007ApJ...668..949B}, \citealt{2008AJ....135.1024P}, \citealt{2012MNRAS.421.2109B}; and numerical simulations: \citealt{2011MNRAS.416.2359B}); and their significant gas reservoirs (e.g. \citealt{2000MNRAS.315..791S}, \citealt{2010PASP..122..683K}).

The interactions of the MCs with each other and with the Milky Way (MW) are believed to have produced several gaseous features: the Magellanic Bridge (MB), first discovered by \cite{1963AuJPh..16..570H}, a common envelope of HI in which both the Large Magellanic Cloud (LMC) and the Small Magellanic Cloud (SMC) are embedded; the Magellanic Stream, a pure-gaseous stream trailing the MCs as they orbit the MW (\citealt{1974ApJ...190..291M}, \citealt{2010ApJ...723.1618N}); and the Leading Arm, connecting both of the MCs with the MW (\citealt{2000PASA...17..1P}). Theoretical works (e.g. \citealt{1994MNRAS.266..567G}; \citealt{2003MNRAS.339.1135Y}; \citealt{2012ApJ...750...36D}) suggest that the MB has a tidal origin. In a recent encounter between the MCs, the LMC stripped large amounts of HI from the SMC to create the MB (e.g. \citealt{2007MNRAS.381L..11M}), possibly `puffing-up' the stellar distribution on the northeastern side of the SMC (\citealt{2011ApJ...733L..10N}). However, only stars younger than 1\,Gyr old, consistent with in-situ formation, have been found to date in the MB's main HI ridge-line  (\citealt{2007ApJ...658..345H}), while only hints of intermediate-age stars are present in the inter-Cloud region [from carbon stars (e.g. \citealt{2000AJ....119.2789K}); giant stars $\sim$6\,kpc from the SMC \citep{2011ApJ...733L..10N}; old stars in the MB from the analysis of 2MASS\footnote{http://irsa.ipac.caltech.edu/applications/2MASS/IM/} and WISE\footnote{http://wise.ssl.berkeley.edu/} near infrared data  \citep{2012arXiv1209.0216B};  and intermediate-age stars in the south at $\sim$6.5\,kpc \citep{2007ApJ...665L..23N}]. 
Recently, \cite{2011ApJ...737...29O} found metallicity differences in 30 red giants located in the outskirts of the LMC, which they interpreted as a strong
evidence that this kinematically distinct population originated in the SMC. This may be evidence of tidally stripped SMC stars, presumably removed in a past LMC-SMC interaction. This could mean that  the ram pressure stripping -- believed to have played an important, or even dominant, role in forming the MB and Stream (\citealt{2005MNRAS.363..509M}) -- would not likely be the main driver of the MB. 

With the purpose of shedding light on the stellar content of the MCs' inter-Cloud area, we have initiated the MAGIC (MAGellanic Inter-Cloud) project, aimed at disentangling the population age and distribution in the inter-Cloud region. In this paper, we present the first results from MAGIC. We performed quantitative studies of the star formation history (SFH) and age-metallicity relations (AMRs) of two large fields in the inter-Cloud region. Our fields are deliberately chosen to lie away from the main HI ridge-line region (around $\sim 1.5$\,kpc in projection on the northern side). There are several reasons for this. 
First, a tidally stripped stellar population is expected to inhabit a broad region around the HI ridge-line of at least the width of the SMC (e.g. \citealt{2006MNRAS.366..429R}; \citealt{2008gady.book.....B}).
 Second, if ram pressure stripping from the MW halo has played a significant role in the evolution of the Magellanic system, then it is possible that the gaseous MB is now systematically displaced from the region where the tidally stripped stars between the MCs lie. Third, no deep photometry has been performed in these fields before (or indeed anywhere around the MB). This is of key importance since reaching the oldest main sequence (MS) turnoffs allows us to break key age-metallicity degeneracies required to detect an intermediate age population (see e.g. \citealt{2008PASP..120.1355N}). 

This paper is organized as follows. In Section \ref{data}, we describe the observations, data reduction and photometry. The analysis and results are discussed in Section \ref{results}. Finally, in Section \ref{discussion} we present our conclusions.

\section{Observations, Data Reduction and Photometry}\label{data}
 
With the aim of characterizing the stellar content of the inter-Cloud region, we obtained B and R band images of 6 fields during the nights of 4$^{\rm th}$ to 7$^{\rm th}$ December 2010 using the WFI on  the 2.2m telescope in La Silla. In order to compensate for the small number statistics in the observed inter-Cloud fields, we merged them into two large fields. 
These fields cover two large areas (0.75$^{\circ}$$\times$0.75$^{\circ}$) in the Bridge region. They are depicted as large red diamonds in  Figure \ref{fig_bridge2} (we will return to this Figure later): B1, located at a distance of $\sim$8.7$^{\circ}$ ($\sim$9 kpc) from the LMC, and B2, at $\sim$6.7$^{\circ}$ ($\sim$8 kpc) from the SMC. In Table~\ref{ObsTable}, we present the centers of the observed fields, the observing conditions, and the 50\% completeness level of the data.

The basic data reduction and astrometric solution was performed by the Cambridge Astronomical Survey Unit pipeline (CASU; \citealt{2001NewAR..45..105I}). 
 For each night, master bias and master flat frames were created from the individual bias and sky flat frames, then master frames from adjacent nights were combined to form a rolling-average frame which was then applied to the data. Each science frame then underwent source extraction and an astrometric solution was determined through crossmatching with the 2MASS point source catalogue. 
The astrometric accuracy is obtained through a Median Absolute Deviation estimator comparing the offsets between the R.A. and DEC of the source versus the 2MASS catalogue. 
This is then scaled, assuming the errors are Gaussian, forming a robust rms estimate of the overall fit.
This resulted in an accuracy of better than 0$\arcsec$.2 in almost all fields and in a few cases better than 0$\arcsec$.1.

The final photometric catalogue was undertaken using {\tt DOLPHOT}\footnote{http://purcell.as.arizona.edu/dolphot/} \citep{2000PASP..112.1383D}.
 {\tt DOLPHOT} extracts the sources and determines the magnitudes for the two bands, B and R, simultaneously. This was run on all of the fields creating a catalogue of sources per 
 field in both filters.
 The output files of {\tt DOLPHOT} include detailed information for each star including the position and global solution such as: $\chi$, sharpness, roundness, and object type. 
 For our purposes, we selected  a value for $\chi$ of $\sim$4, reasonable for good stars within relatively crowded fields, a value of (sharpness)$^{2}$$\lesssim$0.1 for a perfectly-fit star, and object type (which ranges from 1 to 5) of 1 indicating a `good star'. These parameters allow us to remove all spurious objects from the photometry, including background galaxies. 
 Typically this corresponds to $\sim$2400 objects removed per field from our area of interest in CMD space which is consistent with predictions of expected background contamination.
 
During the photometric nights of  4th December 2010 and  6th December 2010 we took short exposures (60 seconds) in the center of B1 and B2 (see Figure \ref{fig_bridge2}) in both, B and R-bands. This
exposures overlap with all the long exposures and were used to provide the final calibration.
Landolt (\citealt{2009AJ....137.4186L}) and Stetson's (\citealt{2000PASP..112..925S}) standard fields SA95 and RU149  
were also observed during photometric nights at different air masses ranging from 2.2 to 1.15.
We obtained several 30 seconds exposures per standard field per filter (in B and R-bands).  
 Since the standard images are uncrowded fields, no profile-fitting photometry was necessary for them.

To put our observations into the standard system, we used the following transformation equations:

\begin{eqnarray} \label{transf}
\nonumber B=b+\alpha_{b}+\beta_{b}(B-R)+\gamma_{b}X_{b} \\
 \\
\nonumber R=r+\alpha_{r}+\beta_{r}(B-R)+\gamma_{r}X_{r}
\end{eqnarray}

 where $(b,r)$ and $(B,R)$ are the instrumental and standard magnitudes respectively, and (X$_{b}$, X$_{r}$) are the airmasses.
No time dependence terms were added since a preliminary fit showed no trends in the residuals of both {\it B} and {\it R} with time.
The color terms ($\beta$$_{b}$, $\beta$$_{r}$) and extinction coefficients ($\gamma$$_{b}$, $\gamma$$_{r}$),
as well as the zero-points ($\alpha$$_{b}$, $\alpha$$_{r}$) are unknown, most likely constant, transformation coefficients and should be calculated.  
Both the color-dependent term and the zero-point in the transformation, are expected to be constant properties of the telescope/filter/detector
combination.
The above equations  were applied to the standard star magnitudes, and solved for
$\alpha$$_{b}$, $\alpha$$_{r}$, $\beta$$_{b}$, $\beta$$_{r}$, $\gamma$$_{b}$ and $\gamma$$_{r}$, for each night, using a custom program.
Then, we followed an iterative procedure to refine our photometric transformation. First,
a new set of unique ($\alpha$$_{b}$, $\alpha$$_{r}$) and ($\beta$$_{b}$, $\beta$$_{r}$) values was obtained by
 imposing the
extinction coefficients ($\gamma$$_{b}$, $\gamma$$_{r}$) corresponding to each night. Then, we applied the resultant zero-points and
color terms to each night and new extinction coefficients were derived for each night. In this way, 
we have a set of ($\alpha$$_{b}$, $\alpha$$_{r}$) and ($\beta$$_{b}$, $\beta$$_{r}$) values 
 and ($\gamma$$_{b}$, $\gamma$$_{r}$) values for each night.

{\tt DOLPHOT} was then subsequently used to perform an artificial star experiment allowing the magnitude completeness of the data to be determined (see \citealt{2007AJ....133.2037N} for details). 
The magnitude completeness was then determined by fitting a logistic function to the ratio of retrieved stars versus input stars.  
Each frame was first processed with {\tt DOLPHOT} to locate the positions of the real stars and then, to ensure good number statistics,  $\sim$48300 stars were introduced per chip one at a time into specific pixel positions. In this manner, {\tt DOLPHOT}  applies a single artificial star in that location and attempts to recover it.
The pixel positions of the artificial stars are chosen such that no two artificial stars probe the same region of the CCD, i.e. there will be no interaction between them, with artificial stars being around 11 pixels apart (roughly PSF radius+ Fitting radius+1) one from another. 
A grid with artificial stars spaced 11 pixels is displaced randomly for each test to assure a dense sampling of the observational effects all over the image of the galaxy. In this way, 
 the artificial stars are spread evenly over the chip and each chip receives the same number of artificial stars. In total, a given exposure has 
 $\sim$384,000 artificial stars inserted into the image.

 The resulting CMDs of field B1 and B2 are shown in Figure \ref{cmds}. As we can see, even at $\sim$8 kpc from the SMC center (field
 B2) the main features of a CMD are still present. We will come back to Figure \ref{cmds} in Section \ref{results}.

 \begin{deluxetable}{lcccclcc}
\tabletypesize{\scriptsize}
\tablecaption{Summary of Observations. The zero-points of the data are 25.25 in B and 24.68 in R. The atmospheric extinction parameters, $\epsilon_{B}=0.271$ and $\epsilon_{R}=0.07$, were taken from \citep{2006astro.ph.11262G} and \citep{1977Msngr..11....7T}, respectively, as this gave the best B-R color in comparison with established catalogues.\label{ObsTable}}
\tablehead{
\colhead{Fieldname}& \colhead{RA (\arcdeg)}&\colhead{DEC (\arcdeg)}&\colhead{Date Observed}&\colhead{Seeing} &\colhead{Filter} &\colhead{Airmass} &\colhead{50\% Completeness}
}
\startdata
0220-7230  &   35.168160 &      -72.49942 &      2010-12-05T05:15:31.760 & 1.53\arcsec &  B\footnote{BB\#B/123\_ESO878} &  1.548 & 24.12 \\
0220-7230  &   35.167300 &      -72.49958 &      2010-12-05T05:59:32.720 & 1.44\arcsec &  R\footnote{BB\#Rc/162\_ESO844} &  1.650 & 22.68 \\
0220-7300  &   35.167970 &      -72.99959 &      2010-12-06T05:26:38.510 & 1.89\arcsec &  B &  1.588 & 23.98 \\
0220-7300  &   35.168490 &      -72.99974 &      2010-12-06T06:11:19.400 & 1.92\arcsec &  R &  1.700 & 22.39 \\
0227-7300  &   36.750970 &      -72.99963 &      2010-12-08T05:24:12.400 & 1.88\arcsec &  B &  1.586 & 23.98 \\
0227-7300  &   36.751540 &      -72.99948 &      2010-12-08T06:08:30.380 & 1.45\arcsec &  R &  1.696 & 23.68 \\
0338-7300  &   54.500650 &      -72.99946 &      2010-12-05T07:51:41.300 & 1.74\arcsec &  B &  1.758 & 23.26 \\
0338-7300  &   54.499860 &      -72.99942 &      2010-12-05T06:34:11.750 & 1.59\arcsec &  R &  1.560 & 21.96 \\
0345-7230  &   56.416990 &      -72.49967 &      2010-12-08T06:46:01.980 & 1.63\arcsec &  B &  1.586 & 24.12 \\
0345-7230  &   56.417650 &      -72.49968 &      2010-12-07T07:03:34.480 & 1.32\arcsec &  R &  1.618 & 22.46 \\
0345-7300  &   56.417620 &      -72.99945 &      2010-12-07T05:12:35.840 & 1.39\arcsec &  B &  1.438 & 23.83 \\ 
0345-7300  &   56.416970 &      -72.99972 &      2010-12-06T07:39:23.810 & 1.77\arcsec &  R &  1.710 & 21.38\\
\enddata	
\end{deluxetable}

\section{Results}\label{results}

\subsection{The star formation history in the inter-Cloud region}
\begin{figure}
 \plotone{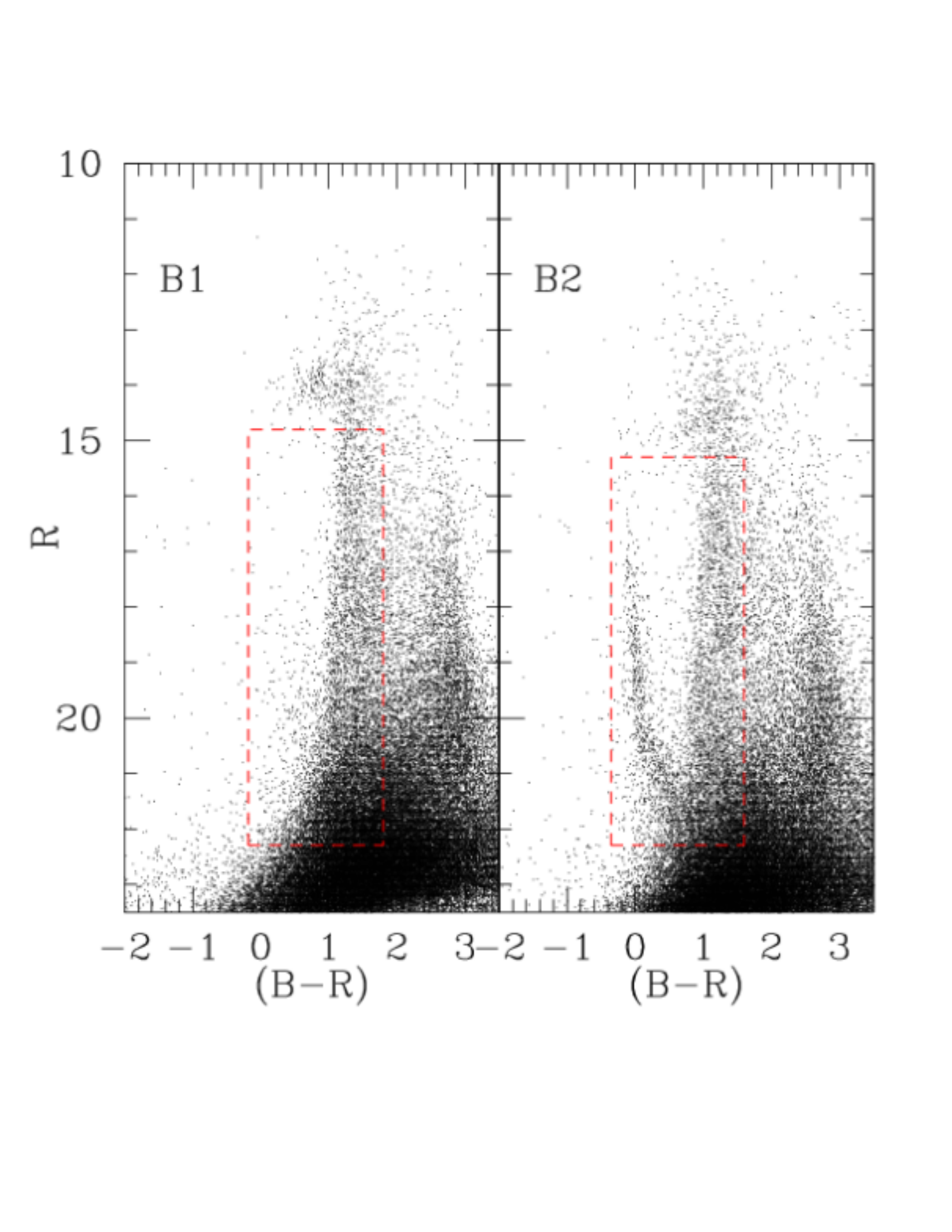}
 \vspace{-45mm}
\caption{CMDs from both of our large fields: B1 (left; closer to the LMC) and B2 (right; closer to the SMC). The spatial location of these two fields is marked by the red diamonds in Figure \ref{fig_bridge2}. The red dashed lines mark the regions used for our CMD fitting analysis.}\label{cmds}
\end{figure}

The CMD fitting technique was carried out using the software package {\tt MATCH} (\citealt{2002MNRAS.332...91D}). The observed CMD was converted into a Hess diagram and compared with synthetic CMDs of model populations from \cite{2004A&A...422..205G}. Theoretical isochrones were convolved with a model of the completeness and photometric accuracy in order to create the synthetic CMDs. We also obtained a foreground estimation based on a Galactic structure model given by {\tt MATCH} that is included in the software as an extra model population. The software uses a maximum-likelihood technique to find the best linear combination of population models, resulting in an estimate of the SFH and the AMR. In order to obtain the SFHs for the inter-Cloud fields analyzed here, we used the color range -0.18 $\leq$ (B-R) $\leq$ 1.8 and the magnitude range R $\leq$ 22.3, B $\leq$ 22.7 for the CMD of B1 and -0.35 $\leq$ (B-R) $\leq$ 1.6 with R $\leq$ 22.3, B $\leq$ 22.7 for the CMD of B2 as marked by the red dashed lines in Figure \ref{cmds}. In this way, we avoid the inclusion of red low mass foreground stars in the fitting. 
To further test the influence of unidentified background galaxies contaminating our sample in the region between 21 $\leq$ R $\leq$ 22.3, we calculated the ratio of objects in the red clump to the number of objects at the lower edge of the CMD (21 $\leq$ R $\leq$ 22.3). For field B2, we find this ratio to be ~15\%, while in the qj0111 SMC wing field from  \cite{2007AJ....133.2037N}, where the contamination is known to be very low, we obtain a ratio of $\sim$11\%. This exercise, together with the constrains given by {\tt DOLPHOT} (see Section \ref{data}), make it clear that any background contamination has been minimized for our dataset.

\begin{figure}[h!]
\begin{center}
\subfigure[SFH and AMR for field B1.]{
\includegraphics[scale=0.35]{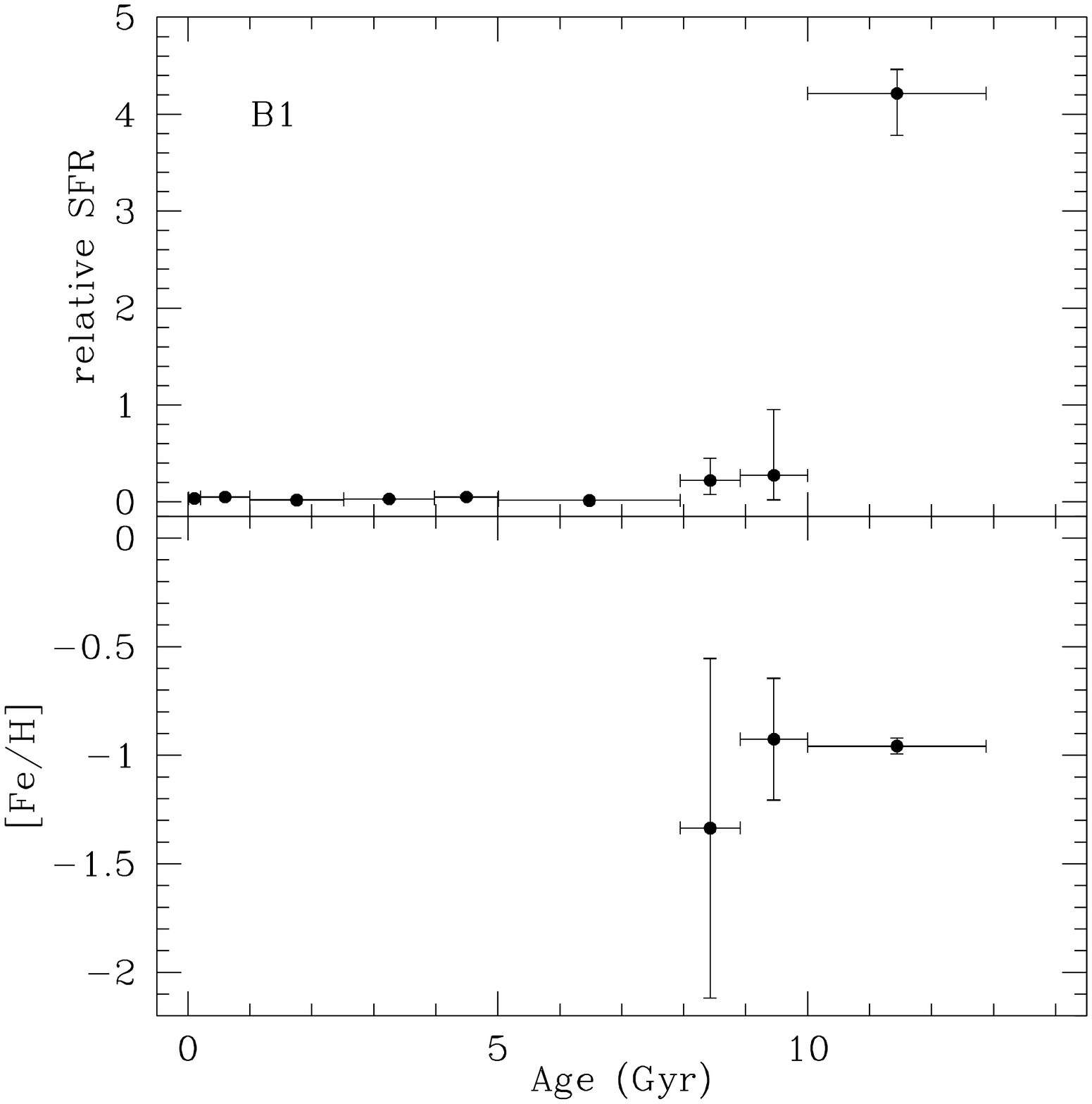}}
\subfigure[SFH and AMR for field B2.]{
\includegraphics[scale=0.35]{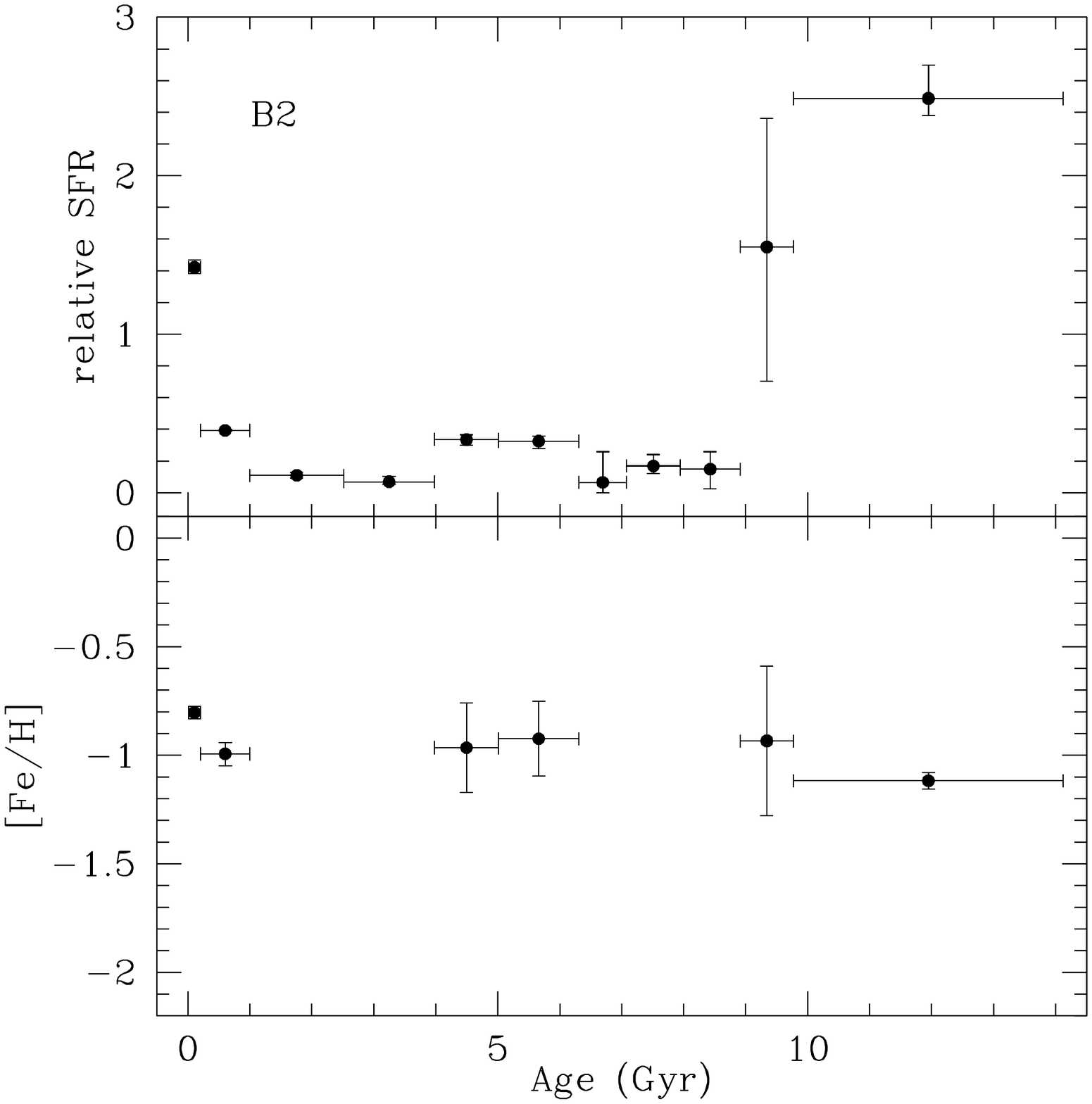}}
\caption{\label{sfhamr} SFHs and AMRs for the CMDs of fields B1 (left) and B2 (right). The average SFRs over the total age range is shown in the upper panels. The vertical error bars give the full range of SFRs found in the fits. The SFR-weighted metallicity (i.e. the AMR) is shown in the lower panels. The vertical error bars give the standard deviation in the metallicities of the fits. Horizontal bars in both the upper and lower panels indicate the width of the age bins.}
\end{center}
\end{figure}

We recovered the SFH and metallicity for both fields B1 and B2 using age bins in the range from 10\,Myr to 13\,Gyr (9 bins for B1 and 11 bins for B2\footnote{We can afford more bins for field B2 since it has more stars.}) and metallicities in the range -2.4$\lesssim$ [Fe/H] $\lesssim$ 0 in bins of width 0.2\,dex. We used a Salpeter initial mass function (IMF) and assumed a 30\% binary fraction\footnote{Since fields B1 and B2 contain predominately intermediate- and old-age stars, the choice of IMF is not critical (see e.g. \citealt{2010ApJ...714..663D}). The binary fraction we assume is taken from \cite{2007AJ....133.2037N}.}. The fitting software then finds a linear combination of model CMDs that best fit the observed CMD. The resulting relative star formation rates (SFRs) and the [Fe/H] as a function of time for B1 and B2 are presented in Figure \ref{sfhamr}. 

The upper panels of Figure \ref{sfhamr} show the SFHs for both fields. The SFR for each bin is the average of the values given by the fits at the different distances and the error bars represent the complete range of SFR values found. The lower panels of Figure \ref{sfhamr} show the AMRs. The metallicities are only plotted for age bins with significant star formation (SFR $\geq$0.2). The metallicity of each bin is the average of the fits weighted by the SFH. Error bars in the AMR represent the standard deviation. Horizontal bars in both the SFH and the AMR plots represent the width of the age bins used. 
 
In both fields, there is a conspicuous old population (older than $\sim$10 Gyr old). Very little star formation took place after then in the inter-Cloud field B1 with only hints of the presence of stars of intermediate-age between $\sim$7-10 Gyr. However, at 6.7$^{\circ}$ ($\sim$8 kpc) from the SMC center, in field B2, there are intermediate-age stars and a young component. In the last 1\,Gyr, field B2 had an increment in the SFR with a conspicuous peak at around 200\,Myr ago (corresponding to the young MS plume in the CMD). 

The AMR of field B1 (Figure \ref{sfhamr}) is consistent with being constant, with [Fe/H]$\sim$-1. The AMR of field B2 is almost constant with [Fe/H]$\sim$-1 until the last few Myr when it goes up to [Fe/H]$\sim$-0.8. These values are in good agreement with the Calcium triplet metallicities obtained for the outskirts of the SMC (\citealt{2008AJ....136.1039C}) and -- within the quoted uncertainties -- the LMC (\citealt{2011AJ....142...61C}). Note that these AMRs obtained from CMD fitting should only be taken as indicative of the metallicity range; accurate metallicities are only possible with spectroscopy.

\begin{figure}[h!]
\begin{center}
\includegraphics[scale=0.5]{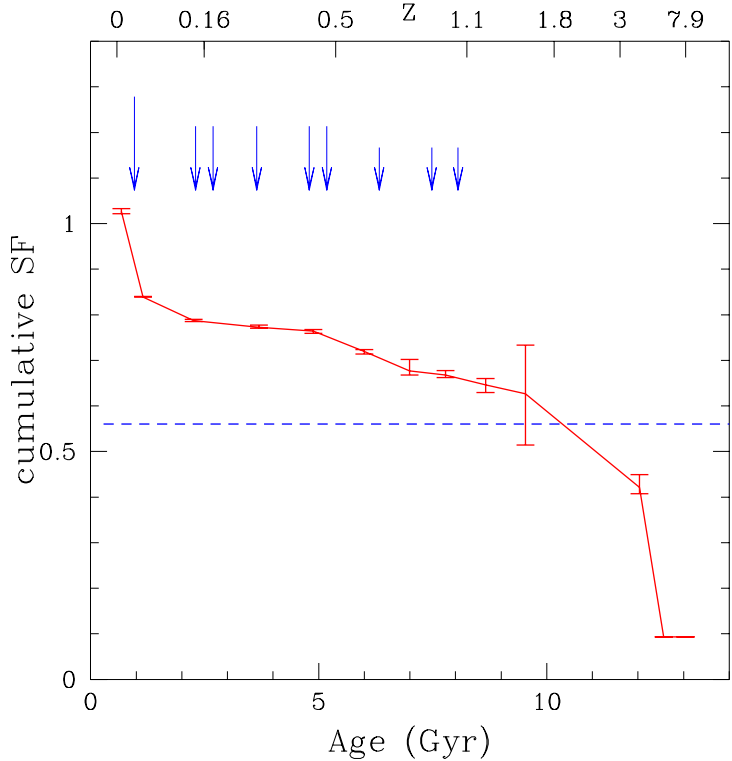}
\vspace{-25mm}
\caption{\label{cumulative} Cumulative SFH of field B2 (see figure \ref{sfhamr}). The blue dashed horizontal line shows when 50\% of the stars were formed. The upper-$y$-axis shows redshift for our current cosmology. The (tentative) pericenter passages that indicate possible past interactions between the LMC and SMC [taken from \cite{2006ApJ...652.1213K}] are marked by blue arrows. The size of the arrows indicate how close the encounter was (larger arrows indicate a closer pericentric passage).}
\end{center}
\end{figure}

Further quantitative information on the stellar distribution is given by the cumulative SFH that provides the fraction of stellar mass formed prior to a given time. This is shown for field B2 in Figure \ref{cumulative}. The upper-$y$-axis shows the redshift as a function of time\footnote{The redshifts were obtained using the following cosmology: $H_{0}$=71, $\Omega$$_{M}$=0.270, $\Omega_\Lambda$=0.730 (http://map.gsfc.nasa.gov/)}. The blue dashed line indicates that 50\% of the stars were formed prior to z$\sim$1.8 ($\sim$10\,Gyr ago). A significant amount ($\sim$28\%) of stars were formed between then and 1\,Gyr ago. The remaining stars formed in the last Gyr with a conspicuous increment (20\% of the stars) in the past few Myr. This recent burst appears to correlate with the most recent pericentric passage between the Clouds ($\sim$\,0.2\,Gyr ago; \citealt{2006ApJ...652.1213K}). The arrows in Figure \ref{cumulative} show this last pericentric passage, together with (tentative) previous close encounters up to $\sim$8\,Gyr ago derived from numerical modeling \citep{2006ApJ...652.1213K}. The size of the arrows represent the proximity of the pericentric passage, with longer arrows representing closer and therefore more extreme interactions. Except for the last encounter $\sim 200$\,Myr ago, when the MCs were $< 5$\,kpc from each other, there seems to be no correlation between enhancements in the SFR and the pericentric passages, in agreement with the previous results from the 12 SMC fields of 
\cite{2009ApJ...705.1260N}. We discuss this further in Section \ref{discussion}.

\begin{figure}
\begin{center}
\includegraphics[angle=0,width=1.02\textwidth]{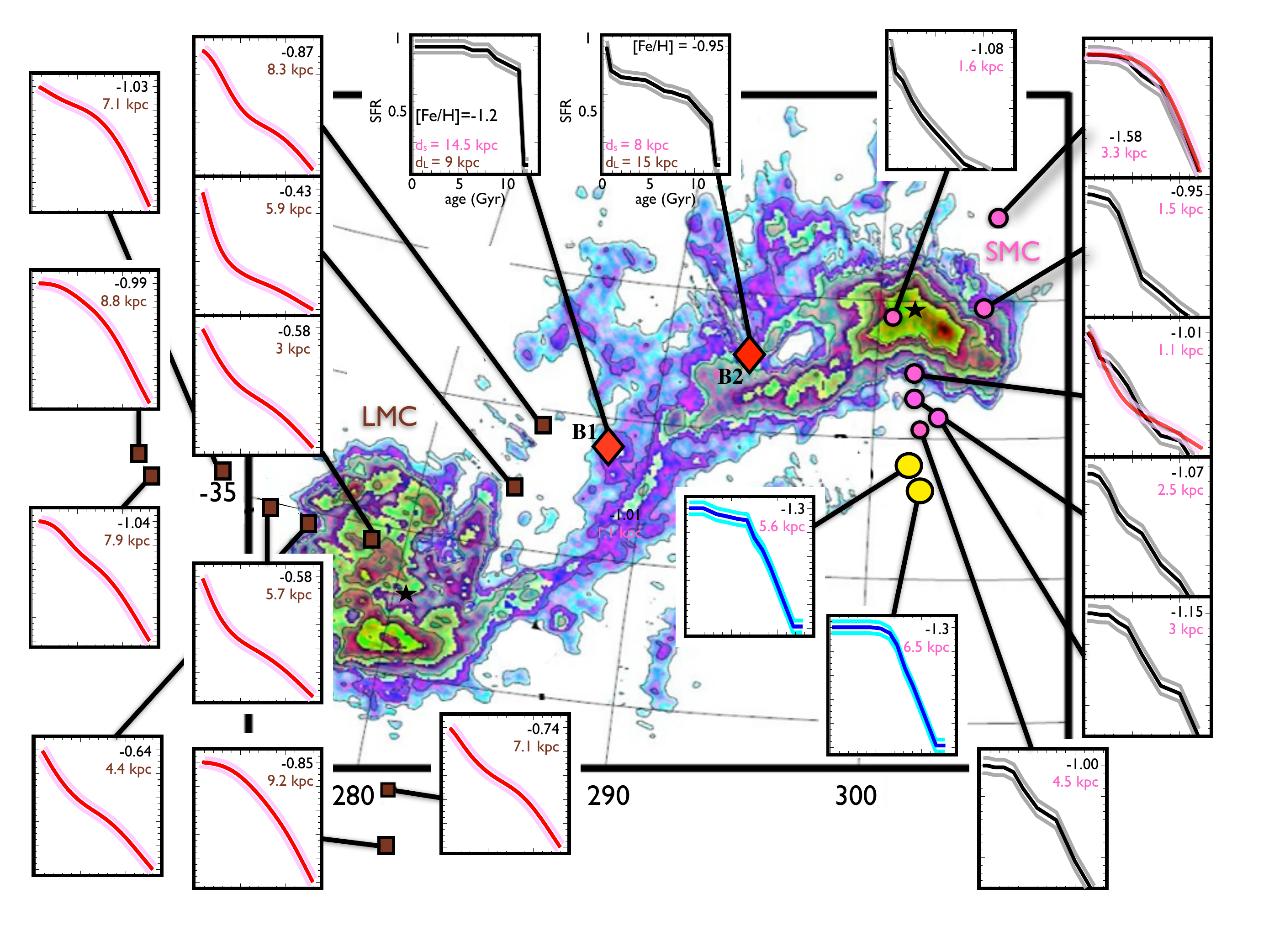}
\vspace{-15mm}
\caption{Spatial location of the two fields presented in this paper (large red diamonds denoted as B1 and B2) together with the SMC fields from \cite{2007AJ....133.2037N} and \cite{2009ApJ...705.1260N} (small magenta circles); \cite{2007ApJ...665L..23N} (large yellow circles); and the LMC fields from \cite{2011AJ....142...61C} (brown squares). The fields are superimposed over the HI map from \cite{1998Natur.394..752P}. Metallicities of the SMC fields are taken from \cite{2008AJ....136.1039C}, except for the yellow circle fields where the metallicity comes from isochrone fitting of \cite{2007ApJ...665L..23N}. Metallicities for the LMC are taken from \cite{2011AJ....142...61C}. Inset figures show the cumulative SFHs (where available; black lines), the age distributions (magenta lines; see text for details) and the stellar content determined from isochrone fitting (blue lines; see text for details). In order to see how the SFHs and the age distributions compare we plot both in two of the SMC fields at 1.1 kpc and 3.3 kpc, where both have been measured. [Fe/H] and distance from the center of the SMC/LMC are also marked on each panel. In the case of B1 and B2, distances to both the LMC (d$_{L}$; brown text) and the SMC (d$_{S}$; magenta text) are marked. Except for fields B1 and B2, axis labels for the inset panels (identical to fields B1 and B2) are omitted for clarity. The center of both the LMC and the SMC are marked with a black filled star.}\label{fig_bridge2}
\end{center}
\end{figure}
  
\subsection{Are these intermediate-age stars tidally stripped?}

We now turn to the issue of whether there are tidally stripped stars in the inter-Cloud region. To determine this, we compare our results for the SFH and AMR of fields B1 and B2 with other fields around the SMC and LMC taken from the literature, as compiled in Figure \ref{fig_bridge2}. The cumulative SFHs (i.e., the average SFR at a given age times the width of the age bins) for the new fields B1 and B2 (large red diamonds) are plotted together the cumulative SFHs for the SMC (magenta and yellow circles) and for the LMC (brown squares) taken from \cite{2007ApJ...665L..23N}, \cite{2007AJ....133.2037N}, \cite{2009ApJ...705.1260N} and \cite{2011AJ....142...61C}. All the fields are superimposed on the HI map of the MCs from \cite{1998Natur.394..752P}. 
  For the LMC fields we obtained the SFHs from the ``age distributions" from \cite{2011AJ....142...61C} (red lines with errors as magenta lines). These age distributions were obtained by comparing the CMD position and spectroscopic (Ca II triplet) metallicity with synthetic CMD models. Since spectra are only available for the brightest stars, this is less accurate than a SFH obtained from a full deep CMD (black lines with errors marked as grey lines). For two fields (see the right-most inset panels in Figure \ref{fig_bridge2}), we show both the quantitative SFHs from CMD fittings and the age distributions; they agree well within the quoted 1-$\sigma$ uncertainties. 
For the two outermost SMC fields in the South (marked as large yellow circles) we obtained the SFH from isochrone fitting. This is a more qualitative approach; however, the depth of the CMDs, reaching the oldest MS turnoffs [see \citet{2007ApJ...665L..23N}] and the little crowding in these outskirt fields allow us to have a very accurate estimation of the SFH. 
  
The metallicities for all the LMC fields are taken from the Calcium triplet study of \cite{2011AJ....142...61C} and the metallicities for the SMC fields for which quantitative SFHs are available (black lines, from the small magenta circles) are from the Calcium triplet of \cite{2008AJ....136.1039C}. The metallicities of the fields presented here (red large diamonds) were obtained through the CMD fitting technique while the ones for the two southernmost SMC fields (large yellow circles) were obtained overlapping isochrones. 

Fields within $\sim$7\,kpc from the LMC center at all position angles have near-continuous star formation with a star formation `burst' at $\lesssim 3$\,Gyrs for the fields closer to the LMC center; more distant fields in the the direction away from the SMC show a slowly declining SFH. By contrast, in the SMC there are few stars formed over the past few Gyr at distances $\gtrsim$2.5\,kpc from the SMC center. Interestingly, fields B1 and B2 resemble SMC fields much more closely than LMC fields. Field B2, located at 8\,kpc from the SMC center, presents both active star formation and intermediate-age stars, with an enhancement in the star formation at $\sim$4\,Gyr, resembling the SMC fields that lie closer than $\sim$2.5\,kpc. Similarly, field B1 -- that shows only stars older than $\sim 7$\,Gyrs -- most closely resembles the distant southern SMC field at 6.5\,kpc, despite lying significantly closer to the LMC than the SMC. The similarity between fields B1 and B2 and other SMC fields that lie much closer in projection to the SMC could be evidence that fields B1 and B2 contain tidally stripped SMC stars. The metallicities of fields B1 and B2 also support this interpretation, though the errors are large. Since the stellar distributions presented in figure \ref{fig_bridge2} are not homogeneous this only allows a qualitative, instead of a more quantitative, interpretation of the tidal scenario. 

As a sanity check of the hypothesis that fields B1 and B2 contain tidally stripped SMC stars, we crudely estimate the projected tidal radius of the SMC by matching field B2 to the SMC fields to the south and west of the SMC. This gives $R_t \sim 1.5 - 2.5$\,kpc. \cite{2006ApJ...652.1213K} estimate that the Clouds passed within $r_p \sim 5$\,kpc of one another $\sim 200$\,Myr ago. The dynamical tidal radius at this pericentric passage is approximately given by (e.g. \citealt{2006MNRAS.366..429R}): 

\begin{equation} 
R_t = \left[\frac{M_\mathrm{SMC}}{M_\mathrm{LMC}}\right]^{1/3} (r_p - R_t)
\end{equation} 
where the masses of the Clouds, $M_\mathrm{SMC,LMC}$, corresponds to that enclosed within $R_t$. Current estimates from the literature suggest that $M_\mathrm{SMC}/M_\mathrm{LMC}(R < 5\,\mathrm{kpc}) \sim 1$ (see e.g. \citealt{2002AJ....124.2639V}; \citealt{1997A&AS..122..507H}), from which we derive $R_t \sim r_p / 2 \sim 2.5$\,kpc, consistent with our value derived above. This is by no means a proof that B2 contains tidally stripped stars; spectroscopy will be required to really confirm or rule out this hypothesis, but it is encouraging.

\section{Discussion and Conclusions}\label{discussion}
 
We have presented the stellar populations of two fields located in the area between the Magellanic Clouds -- the `inter-Cloud' region. The fields -- located at a distance of $\sim$8.7$^{\circ}$ from the LMC (9\,kpc; field B1), and at $\sim$6.7$^{\circ}$ from the SMC (8\,kpc; field B2) -- were chosen deliberately to avoid the MB formed by the main ridge-line of HI gas. We found that the stellar populations and mean metallicities of the fields are similar to fields to the south and west of the SMC but at smaller radii from the SMC center ($\sim 6$\,kpc in the case of field B1 and $\sim 2.5$\,kpc in the case of field B2). The similarity between fields B1 and B2 and other SMC fields that lie much closer in projection to the SMC could be an indication that fields B1 and B2 contain tidally stripped SMC stars.  

Our key result is that intermediate-age stars are present in the inter-Cloud region. We found, through a quantitative CMD fitting analysis, that some $\sim$28\% of the stars in field B2 have intermediate age. These findings are consistent with several independent results in the literature. Carbon stars (unequivocal indicators of intermediate-age population) were already discovered in the inter-Cloud area over a decade ago (e.g. \citealt{2000AJ....119.2789K}); the kinematically peculiar stars found in the LMC by \cite{2011ApJ...737...29O} could be evidence of SMC stars being captured by the LMC; \cite{2011ApJ...733L..10N} found a break in the radial density profile from $\sim$7.5$^{\circ}$ up to $\sim$10.6$^{\circ}$ which could be a population of extra-tidal stars or a bound stellar halo; finally, $N$-body models from \cite{2012ApJ...750...36D} predict that diffuse tidal features are stripped away from the MCs during the interaction between the Clouds that forms the MB. Our findings are at odds, however, with \cite{2007ApJ...658..345H} who found only old stars east of $\alpha$$\sim$03h 18' and $\delta$$\sim$-74$^{\circ}$ and only young stars in their inter-Cloud fields along the MB's main ridge-line, with no evidence of an intermediate-age component. Part of these differences are likely due to the fact that we have deeper CMDs that reach the oldest MS turnoffs, of key importance for finding intermediate-age stars (see e.g. \citealt{2008PASP..120.1355N}).

In addition to the intermediate-age stars, field B2 shows a significant young population that correlates with the last close pericentric passage between the Clouds $\sim$200\,Myr ago, as determined from their proper motions and radial velocities (\citealt{2006ApJ...652.1213K}). This is the only clear starburst associated with a pericentric passage, which could reflect the much closer proximity of the last fly-by (\citealt{2006ApJ...652.1213K}), the difficulty of determining orbits backwards in time \citep{2010MNRAS.406.2312L}, and/or the poorer temporal resolution of the star formation history backwards in time \citep{2012ApJ...751...60D}.

\acknowledgments
We would like to thank the anonymous referee for useful suggestions that help improving the manuscript. 
We also thank Nicolas Martin and Matteo Monelli for their helpful insights during the initial stages of the project. N.E.D.N and B.C.C. would like to thank the Max-Planck Institute for Astronomy. N.E.D.N. and J.I.R. thank the ETH 
Z\"urich. B.C.C. also thanks the Alexander von Humboldt Foundation Fellowship under which funding for this project was completed. R.C. acknowledges the funds from the Spanish Ministry of Science and Innovation under the Juan de la Cierva fellowship. J.I.R. would like to acknowledge support from SNF grant PP00P2\_128540/1.

\end{document}